\documentclass[onecolumn,tighten,times]{aastex631}
\usepackage{amsmath}
\usepackage{enumitem}

\received{:12 20 2024}
\revised{:28 02 2025}
\accepted{:02 03 2025}
\submitjournal{ApJL}

\shorttitle{A recent SMBHB in the GC}
\shortauthors{Cao et al.}

\begin{document}

\title{A Recent Supermassive Black Hole Binary in the Galactic Center Unveiled  
by the Hypervelocity Stars}

\author[0000-0002-4211-9523]{Chunyang Cao}
\affiliation{Department of Astronomy, School of Physics, Peking University, 
 Beijing 100871, People's Republic of China}

 \author[0000-0002-5310-3084]{F.K. Liu}
\affiliation{Department of Astronomy, School of Physics, Peking University, 
 Beijing 100871, People's Republic of China}
 \affiliation{Kavli Institute for Astronomy and Astrophysics, Peking University, 
Beijing 100871, People's Republic of China}

\author[0000-0001-6530-0424]{Shuo Li}
\affiliation{National Astronomical Observatories, Chinese Academy of Sciences, 
Beijing 100012, People's Republic of China}

\author[0000-0003-3950-9317]{Xian Chen}
\affiliation{Department of Astronomy, School of Physics, Peking University, 
 Beijing 100871, People's Republic of China}
\affiliation{Kavli Institute for Astronomy and Astrophysics, Peking University, 
 Beijing 100871, People's Republic of China}

\author[0000-0002-7237-3856]{Ke Wang}
\affiliation{Kavli Institute for Astronomy and Astrophysics, Peking University, 
Beijing 100871, People's Republic of China}

\correspondingauthor{F.K. Liu}
\email{fkliu@pku.edu.cn}

\begin{abstract}
When a binary of early-type stars from the young stellar populations 
in the Galactic center (GC) region is scattered to 
the vicinity of the supermassive black hole (SMBH) Sgr~$\rm{A}^{*}$, 
one of the components would be tidally ejected as an early-type hypervelocity star (HVS) 
and the counterpart would be captured on a tight orbit around Sgr~$\rm{A}^{*}$. 
Dozens of B-type HVSs moving faster than the Galactic escape speed 
have been discovered in the Galactic halo and are produced most likely by 
the SMBH Sgr~$\rm{A}^{*}$. 
However, the velocity distribution and 
in particular the deficit of the HVSs above $700\, \rm{km\, s^{-1}}$ 
is seriously inconsistent with the expectations of the present models. 
Here we show that the high-velocity deficit is due to the deficiency 
in close interactions of stars with the SMBH Sgr~$\rm{A}^{*}$, 
because an orbiting intermediate-mass black hole (IMBH) of about 15,000 Solar mass 
kicked away slowly approaching stars 50--250 million years ago. 
The SMBH-IMBH binary formed probably after the merger of the Milky Way with the 
Gaia-Sausage-Enceladus dwarf galaxy, 
and coalesced about 10 million years ago, 
leading to a gravitational recoil of Sgr~$\rm{A}^{*}$ 
at a velocity of 0.3--0.5$\, \rm{km\, s^{-1}}$ 
and to a change of the HVS ejection scenarios. 
The SMBH-IMBH binary scenario predicts the formation of the S-star cluster at the GC 
with the distribution of the orbital size and stellar ages 
that are well consistent with the observations.

\end{abstract}

\section{Introduction} \label{sec:introduction}
Recent observations have revealed that a few percent of stars 
of the nuclear star cluster (NSC) are young and formed over the past $500\, \rm{Myr}$ 
with an interruption at $80$--$150\, \rm{Myr}$ ago \citep{schodel_milky_2020}. 
Another population of young stars of ages about $6\, \rm{Myr}$ is also detected 
as the clockwise and counterclockwise disks (hereafter CWD) in the Galactic center 
\citep[GC;][]{levin_stellar_2003,paumard_two_2006,fellenberg_young_2022,jia_stellar_2023}.

Most early-type stars of the young stellar populations 
are born in binary or multiplicity \citep{offner_origin_2022,gautam_estimate_2024}. 
When an early-type binary star is scattered to the vicinity of 
the supermassive black hole (SMBH) Sgr~$\rm{A}^{*}$,
it could be tidally separated, 
resulting in the ejection of an early-type hypervelocity star (HVS) 
\footnote{We refer to HVSs as ejected stars with SMBH origins and unbound velocities 
following \cite{brown_hypervelocity_2015}.} 
and the capture of the counterpart of the same age on a 
tightly bound orbit around Sgr~$\rm{A}^{*}$ 
\citep{hills_hypervelocity_1988,gould_sagittarius_2003}. 
HVSs could also be produced by the gravitational slingshot effect 
when a supermassive black hole binary (SMBHB) is present at the GC \citep{yu_ejection_2003}. 
SMBHBs may form at galaxy mergers and candidates in normal galaxies have been discovered 
\citep[e.g.,][]{liu_milliparsec_2014}. 
The Gaia and large ground-based spectroscopic surveys have revealed that the 
Milky Way (MW) has merged successively with 9--10 dwarf galaxies in the past 
\citep{naidu_evidence_2020},  
leaving no bound star around Sgr~$\rm{A}^{*}$. 
Since dwarf galaxies with total stellar mass $\leq 10^{9.5}M_{\odot}$ are expected to host   
intermediate-mass black holes (IMBHs) of mass $10^{3}$--$10^{6}M_{\odot}$ 
at their centers 
\citep[e.g.,][]{reines_relations_2015,greene_intermediatemass_2020,sanchez_deeper_2024}, 
it is possible that some of the mergers brought in IMBHs 
and an SMBH-IMBH binary once formed at the GC \citep{begelman_massive_1980}.

About 20 late B-type HVSs have been identified in the Galactic halo 
\footnote{The only exception is S5-HVS1 \citep{koposov_discovery_2020}, 
and we will briefly discuss its origin in Section~\ref{sec:dis}.}
with velocities of $300$--$700\, \rm{km\, s^{-1}}$ in the Galactic rest frame     
and are ejected from the GC about $50$--$250\, \rm{Myr}$ ago
\citep{brown_mmt_2014,brown_gaia_2018a}. 
The late B-spectral type suggests that 
their progenitors should have formed in the recent $600\, \rm{Myr}$ 
and most likely originate in the young stellar population of the NSC. 
However, no expected counterpart of age between 50 and 250 Myr is detected 
in the S-star cluster \citep{habibi_twelve_2017}. 
Also, the observed velocity distribution of the halo HVSs 
is severely at odds with the expectations of the existing models. 
Particularly, no halo HVS is found above $700\, \rm{km\, s^{-1}}$, 
while about $40\%$--$60\%$ HVSs are expected to have velocities 
$\gtrsim 700\, \rm{km\, s^{-1}}$, 
produced either through the tidal separations of the compact stellar binaries 
of semimajor axis $\lesssim 0.3\, \rm{au}$ 
\citep{zhang_spatial_2010,rossi_velocity_2014,generozov_constraints_2022} or 
through the strong slingshot ejections of stars by the SMBHB 
in the full-loss-cone regime 
as the interaction loss cone being efficiently populated over time 
\citep{levin_ejection_2006,sesana_interaction_2008,zhang_spatial_2010,
darbha_gravitational_2019,evans_constraints_2023}. 
The truncation semimajor axis $0.3\, \rm{au}$ is approximately an order of magnitude 
larger than the separation of a contact binary and inconsistent 
with the observations of the early-type binaries \citep[e.g.,][]{moe_mind_2017}.


Here we propose a model in which the late B-type halo HVSs are produced by 
an evolving SMBHB at the GC interacting with 
the young stellar populations of the NSC 
through both the binary tidal separations and the slingshot ejections. 
We show that the deficiency of the halo HVSs above $700\, \rm{km\, s^{-1}}$ 
is due to the blocking effect of the secondary IMBH. 
The SMBHB has a semimajor axis of 
$a_{\rm{MBHB}}\approx 100\, \rm{au}$ when the halo HVSs are ejected. 
Because of the large orbit size of the SMBHB, 
wandering stars would gradually approach and have repeated slingshot interactions with 
the secondary IMBH (empty-loss-cone regime) to gain energy 
and be ejected away before deeply penetrating into the orbit of the SMBHB. 
The very center region around the primary SMBH is thus empty of orbits and 
few binary tidal separations could occur therein. 
The blocking effect would effectively result in 
the deficit of the tidal separations of the compact binaries 
with tidal separation radius $r_{\rm{bt}}\ll a_{\rm{MBHB}}$ and 
thus of the HVSs with velocities above the associated specific value.

This Letter is organized as follows. 
We introduce our model for HVS production in Section~\ref{sec:model}. 
Section~\ref{sec:num_methods} details the procedures to 
get the mock catalog of HVSs and captured stars from our model 
and to constrain the model parameters. 
The results are presented in Section~\ref{sec:results}. 
We summarize our findings and 
make brief discussions in Section \ref{sec:dis}.

\section{Model} \label{sec:model}

\subsection{Production of Hypervelocity Stars by 
an Evolving Supermassive Black Hole Binary} \label{subsec:evolution}

We consider an SMBHB consisting of 
the SMBH Sgr~A$^{*}$ of mass $M_{1}=4\times 10^{6}M_{\odot}$ \citep{reid_proper_2004} 
and a secondary IMBH of mass $M_{2}=q_{\rm{MBHB}}M_{1}$ with $q_{\rm{MBHB}}$ the mass ratio.  
The SMBHB is embedded in the NSC, 
whose density profile is modeled as 
\begin{equation} \label{equ:NSC_density}
    \rho(r)=\rho(r_{\rm{b}})2^{(\beta-\gamma)/\alpha}
    \left(\frac{r}{r_{\rm{b}}}\right)^{-\gamma}
    \left[1+\left(\frac{r}{r_{\rm{b}}}\right)^{-\alpha}\right]
    ^{(\gamma-\beta)/\alpha},
\end{equation}
where $(\alpha, \gamma, \beta)=(10,1.13,3.5)$, 
$r_{\rm{b}}=3.1\, \rm{pc}$ is about the influence radius of the Sgr~A$^{*}$ 
\citep{schodel_distribution_2018}, 
and $\rho(r_{\rm{b}})$ is calibrated so that the enclosed stellar mass at 
$r_{\rm{b}}$ equals to $2M_{1}$ \citep{merritt_single_2004}. 
The SMBHB initially evolves under dynamical friction, 
which brings the IMBH to a radius where the enclosed stellar mass 
is comparable to $M_{2}$ \citep{matsubayashi_orbital_2007}. 
After that, dynamical friction becomes less efficient 
and the SMBHB evolution is mainly driven by the slingshot ejections of bound stars 
around Sgr~A$^{*}$. 
When most bound stars are depleted, 
the SMBHB evolution slows down. 
This typically happens when the SMBHB becomes hard at the hardening radius 
$a_{\rm{h}}=[4(3-\gamma)]^{-1}q_{\rm{MBHB}}(1+q_{\rm{MBHB}})^{-1}r_{\rm{b}}$ 
\citep{sesana_self_2010}. 
In this study, we consider the production of HVSs by a hard SMBHB, 
because a soft SMBHB with $a_{\rm{MBHB}}\geq a_{\rm{h}}$ 
would evolve quickly and eject HVSs in a burst lasting $1$--$10\, \rm{Myr}$ 
\citep[e.g.,][]{levin_ejection_2006,sesana_interaction_2008}, 
which contradicts with the observed continuous ejection over 
$200\, \rm{Myr}$ \citep{brown_mmt_2014}. 
The evolution of a hard SMBHB to the coalescence is mainly driven by 
losing energy ($E_{\rm{MBHB}}$) and angular momentum ($L_{\rm{MBHB}}$) 
through the slingshot ejections of the distant stars 
that are two-body scattered to the SMBHB in the early stage 
and gravitational wave (GW) radiation in the late stage \citep{begelman_massive_1980},    
described by 
\begin{align} \label{equ:SMBHB_decay}
    \frac{\rm{d}E_{\rm{MBHB}}}{\rm{d}t} &= 
    \left\{\frac{\rm{d}E_{\rm{MBHB}}}{\rm{d}t}\right\}_{\rm{GW}} + 
    \langle C \rangle\frac{2 \langle m_{*} \rangle E_{\rm{MBHB}}}
    {M_{1}(1+q_{\rm{MBHB}})}\Gamma_{\rm{s}}  \nonumber \\
    \frac{\rm{d}L_{\rm{MBHB}}}{\rm{d}t} &= 
    \left\{\frac{\rm{d}L_{\rm{MBHB}}}{\rm{d}t}\right\}_{\rm{GW}}
    -\langle B \rangle\frac{\langle m_{*} \rangle L_{\rm{MBHB}}}
    {M_{1}(1+q_{\rm{MBHB}})}\Gamma_{\rm{s}}. 
\end{align} 
The first (curly bracketed) and the second terms account for, respectively, 
the contribution from the GW radiation \citep{peters_gravitational_1964} 
and the slingshot effect, 
where $\Gamma_{\rm{s}}$ is the slingshot rate 
calculated with the loss-cone theory 
(see Section~\ref{subsec:loss_cone}), 
$\langle m_{*} \rangle \approx 0.3M_{\odot}$ is the averaged stellar mass, 
and $\langle C \rangle$ and $\langle B \rangle$ are, respectively,  
the averaged fractional exchange of specific energy and angular momentum 
during each slingshot ejection \citep{quinlan_dynamical_1996}, 
whose values in the empty-loss-cone regime are obtained with numerical 
scattering experiments (see Appendix~\ref{appendix_sec:harden_rate}).

As the hard SMBHB evolves at the GC, 
it constantly interacts with the stars of the NSC and 
ejects HVSs via two mechanisms. 
The first one is the tidal separations of stellar binaries \citep{hills_hypervelocity_1988}. 
For a stellar binary of semimajor axis $a_{*}$ and total mass $m_{1}+m_{2}$, 
the separation requires its orbital pericenter 
\begin{equation} \label{equ:rbt}
    r_{\rm{p}}\lesssim r_{\rm{bt}}\equiv a_{*} \left(\frac{M_{1}}{m_{1}+m_{2}}\right)^{1/3} 
    \approx 8.7\, \rm{au}  \left({a_{*} \over 0.1 
        \, {\rm au}}\right)  
        \left({m_{1}+m_{2} \over 6 M_\odot} \right)^{-1/3}
        \left({M_{1} \over 4\times 10^6 M_\odot}\right)^{1/3}    
\end{equation}  
and leads to the ejection of one star (say, the star of mass $m_{1}$) with a velocity of 
\citep{hills_hypervelocity_1988,bromley_hypervelocity_2006}
\begin{align} \label{equ:v_eje_bt}
    v_{\rm{e}} \simeq 
    1300\, \rm{km\, s^{-1}}
    \left(\frac{a_{*}}{0.1\, \rm{au}}\right)^{-1/2}
    \left(\frac{2m_{2}}{m_{1}+m_{2}}\right)^{1/2}
    \left(\frac{m_{1}+m_{2}}{6M_{\odot}}\right)^{1/3}
    \left(\frac{M_{1}}{4\times 10^{6} M_{\odot}}\right)^{1/6}. 
\end{align}
The second mechanism is the gravitational slingshot by the SMBHB, 
which could eject a single/binary star of $r_{\rm{p}}\lesssim a_{\rm{MBHB}}$ with 
a typical velocity of \citep{yu_ejection_2003}
\begin{align} \label{equ:v_eje_gs}
    v_{\rm{e}} \simeq
    1000\ \rm{km\ s^{-1}} 
    \left(\frac{a_{\rm{MBHB}}}{100\, \rm{au}}\right)^{-1/2} 
    \left(\frac{q_{\rm{MBHB}}}{0.01}\right)^{1/2}
    \left(\frac{M_{1}}{4\times 10^{6} M_{\odot}}\right)^{1/2}.
\end{align}

\subsection{Interaction Rate of Stars/Binaries with 
the Supermassive Black Hole (Binary)} \label{subsec:loss_cone}

As introduced above, both HVS ejection mechanisms require 
stars/binaries to come close to the SMBH(B) and interact with it. 
All the interacting stars with $r_{\rm{p}}\leq r_{\rm{lc}}$ occupy a region 
$L\leq L_{\rm{lc}} \approx (2GM_{1}r_{\rm{lc}})^{1/2}$
in the angular momentum ($L$) space called loss cone. 
Provided that a fraction $\Theta $ of stars at radius $r$ 
get into the loss cone during dynamical timescale $T_{\rm{d}}$, 
the interaction rate can be expressed as 
\begin{equation} \label{equ:Rate_old}
    \Gamma = \int
    \frac{\Theta(r)}{T_{\rm{d}}(r)}
    \frac{4\pi r^{2} \rho(r)}{\langle m_{*} \rangle}
    \rm{d}r. 
\end{equation}
The loss-cone theory 
\citep{frank_effects_1976,lightman_distribution_1977,cohn_stellar_1978}
suggests that $\Theta $ depends on $L_{\rm{d}}$, 
the evolution of the stellar angular momentum per orbital period 
induced by two-body scattering. 
If $L_{\rm{d}}\gg L_{\rm{lc}}$, 
the loss cone is in the full regime and $\Theta = L_{\rm{lc}}^{2}/L_{\rm{c}}^{2}$. 
Otherwise, the loss cone is in the empty regime and 
$\Theta = (L_{\rm{d}}^{2}/L_{\rm{c}}^{2}) \ln^{-1}(L_{\rm{c}}^{2}/L_{\rm{lc}}^{2})$. 
Numerically, $L_{\rm{d}}$ can be simulated with \citep{risken_fokkerplanck_1996}
\begin{equation} \label{equ:L_kick}
    L_{\rm{d}} = \frac{L_{\rm{c}}^{2}T_{\rm{d}}}{2T_{\rm{r}}L} + 
    \xi \left(\frac{T_{\rm{d}}}{T_{\rm{r}}}\right)^{1/2}L_{\rm{c}}, 
\end{equation}
where $\xi$ is a random variable following a normal distribution of unit variance, 
$L_{\rm{c}}$ is the angular momentum for a circular orbit, 
and $T_{\rm{r}}$ is the two-body relaxation timescale. 
The first and the second terms represent, 
respectively, a steady drift and a random walk in the angular momentum space. 
We measure the strength of the two-body scattering with a free parameter $\kappa$, 
such that 
\begin{equation} \label{equ:T_r}
  T_{\rm{r}}=\kappa T_{0}=\kappa \frac{\sqrt{2}\sigma^{3}}
  {\pi G^{2}\langle m_{*}\rangle \rho \ln{\Lambda}},
\end{equation}  
where $T_{0}$ provides a fiducial estimate of the two-body relaxation timescale 
\citep{spitzer_evaporation_1958}, 
$\ln{\Lambda} = 15$ is the Coulomb logarithm, 
and $\sigma$ is the one-dimensional velocity dispersion.

The traditional loss-cone theory introduced above 
assumes a quasi-equilibrium system and thus applies to the old stars of the NSC. 
However, the late B-type halo HVSs should be younger than about $600\, \rm{Myr}$. 
The best candidates for their origin would be the 
young stellar population recently revealed in the NSC, 
including about $4\%$ mass fraction of stars 
formed 150--500$\, \rm{Myr}$ ago (hereafter 500$\, $Myr population) 
and another about $0.5\%$ formed over the recent $80\, \rm{Myr}$ 
\cite[hereafter 80$\, \rm{Myr}$ population;][]{schodel_milky_2020}. 
Their orbits are not relaxed yet considering that 
the relaxation timescale in the NSC is much longer than $500\, \rm{Myr}$ 
\citep[e.g.,][]{merritt_distribution_2010}, 
so we calculate their interaction rate by 
applying a modification function based on Equation~(\ref{equ:Rate_old}). 
Here we consider a scenario in which the number density of young stars 
first increases as more stars form during the star formation episode, 
and then decreases with time after the star formation ceases 
and the young stellar population gradually gets relaxed. 
We model the decrease as an exponential decay with timescale $\tau$. 
Provided that the young stars form at a constant rate from 
lookback time $T_{\rm{s}}$ to $T_{\rm{e}}$,
the modification function at lookback time $t$ would be 
\begin{align} \label{equ:t_escape}
    F(t,T_{\rm{s}},T_{\rm{e}}) &=
    \frac{\int^{T_{\rm{s}}}_{\max(T_{\rm{e}},t)} 
    \rm{exp}\left(-\frac{t_{\rm{s}}-t}{\tau}\right)\rm{d}t_{\rm{s}}}
    {\int^{T_{\rm{s}}}_{\max(T_{\rm{e}},t)} 
    \rm{d}t_{\rm{s}}},\ t\leq T_{\rm s}.
\end{align} 
Considering that the mass fractions ($4\%$ and $0.5\%$) are measured today, 
we scale the modification function to the present value 
and calculate the interaction rates for the 500$\, $Myr and 80$\, $Myr populations with 
\begin{align} \label{equ:Rate_t}
    \Gamma_{500}(t) =&\ \ 4\%A_{\rm{e}}\frac{F_{500}(t)}{F_{500}(t=0)} \Gamma \nonumber \\ 
    \Gamma_{80}(t)  =& 0.5\%A_{\rm{e}}\frac{F_{80}(t)}{F_{80}(t=0)} \Gamma, 
\end{align}
where $A_{\rm{e}}$ is a dimensionless parameter 
accounting for the magnitude of the modification. 
We assume that the two populations 
share the same $\tau$ and $A_{\rm{e}}$ for simplicity.

At the late stage, 
the dynamical evolution of the SMBHB would decouple from the ambient stars 
when the slingshot loss cone ($L_{\rm{ss}}^{2}\approx 2G M_{1} a_{\rm{MBHB}}$) 
decays faster than the angular momentum diffusion of the stellar orbit 
due to two-body scattering, i.e.,  
$\left\lvert \mathrm{d}L_{\rm{ss}}^{2}/\rm{d}t \right\rvert  
\geq L_{\rm{d}}^{2}/T_{\rm{d}}$.
After the decoupling, 
the production of HVSs by the slingshot ejections would stop. 
The decouple (rather than the coalescence) of the SMBHB 
marks a transition of the HVS ejection scenario.

\section{Numerical Methods} \label{sec:num_methods}

\subsection{Mock Catalogs of Hypervelocity Stars and Captured Stars} \label{subsec:mock_catalog}

Given the model introduced above and the parameters (see Section~{\ref{subsec:fitting}}), 
we can obtain mock catalogs of HVSs and captured stars. 
To begin with, 
we generate a mock catalog of early-type (O and B) stars of mass $\geq 2.5M_{\odot}$ 
for the 500$\, $Myr and 80$\, $Myr populations with Monte Carlo sampling, 
assuming a Kroupa initial mass function \citep[IMF;][]{kroupa_variation_2001} and 
a constant star formation rate during each star formation episode. 
Some stars are born in binary. 
We set the initial binary fraction as an increasing function of the stellar mass,
which is about 50\% for late B stars and nearly 100\% for O stars 
\citep{offner_origin_2022}. 
We sample the binary semimajor axis, eccentricity, and mass ratio 
according to Section 9 in \cite{moe_mind_2017}. 
We remove the contact binaries whose pericenter distances are 
smaller than the Roche radius \citep{eggleton_approximations_1983}. 
And following \cite{generozov_constraints_2022}, 
the mock binary catalog is further refined by requiring the binaries 
to be more compact (measured by self-binding energy) than a binary 
with critical semimajor axis $a_{\rm{c}}=15\, \rm{au}$ and 
$m_{1}=m_{2}=10M_{\odot}$. 
The refined binary fraction is $f_{\rm{b}}\approx 34\%$.

Next, we work out the evolution of the SMBHB and 
the interaction rate of the young stellar population as functions of lookback time 
(see Section~\ref{sec:model}). 
Accordingly, we sample a certain amount of stars/binaries from the mock catalog 
introduced above and 
numerically simulate their orbital evolution and interaction with 
an evolving SMBHB (before decouple) or a SMBH (after decouple). 
Technically, this would involve three gravitational few-body systems: 
(A) an SMBHB and a star; 
(B) an SMBH and a stellar binary; 
and (C) an SMBHB and a stellar binary. 
Here we detail the simulation of the last system. 
The procedures for the other two systems are similar.

The stellar binary is initially placed on a randomly sampled orbit around the SMBHB 
according to Equation~(\ref{equ:Rate_old}) 
with a pericenter distance of $r_{\rm{p}}=10\times \max(r_{\rm{bt}},a_{\rm{MBHB}})$. 
During each orbital period, 
it would have a close gravitational encounter with the SMBHB at the pericenter, 
whose outcome is obtained by interpolating from scattering experiments  
(see Appendix~\ref{appendix_sec:scattering}). 
Meanwhile, its angular momentum ($L$) would evolve due to two-body scattering, 
which is simulated by performing an isotropic kick 
of $\Delta{L}=L_{\rm{d}}$ (see Equation~({\ref{equ:L_kick}})) 
at the apocenter \citep[e.g.,][]{bradnick_stellar_2017} . 
The simulation lasts for 400 orbital periods and is terminated when: 
(1) the binary gains enough energy from the slingshot effect and 
becomes unbound to the MW, 
or $\Delta{L}$ becomes much larger than $L_{\rm lc}$; 
(2) the binary is tidally separated; 
(3) the two stars collide with each other; 
or (4) one star is tidally disrupted (TDE).

For each ejected HVS, 
we derive its velocity at Galactocentric distance $R_{0}=10\, \rm{pc}$ 
in the gravitational potential of the SMBH Sgr~$\rm{A}^{*}$ and the NSC 
assuming an isotropic ejection. 
Its subsequent propagation in the MW is tracked to the present 
using the program \texttt{Gala} \citep{gala} 
with a fixed time resolution of $0.04\, \rm{Myr}$. 
We use the \texttt{MWPotential2014} model to set the MW potential, which comprises 
a power-law density profile with an exponential cut-off for the bulge, 
a Miyamoto–Nagai model \citep{miyamoto_threedimensional_1975} for the disk, 
and a Navarro–Frenk–White model \citep{navarro_universal_1997} 
for the dark-matter halo
\citep[][and see explicit model parameters therein]{bovy_galpy_2015}. 
The HVS is discarded if 
it evolves off the main sequence (MS) during the journey, 
where the MS lifetime is calculated analytically according to 
\cite{hurley_comprehensive_2000}, 
assuming a metallicity of $Z=2Z_{\odot}$.

For a captured star, 
we calculate the semimajor axis ($a_{\rm{S}}$) of its orbit 
around the SMBH Sgr~$\rm{A}^{*}$. 
We only consider captured stars more massive than $3.5M_{\odot}$, 
a typical lower mass limit for the B-type S-stars 
\citep{eisenhauer_sinfoni_2005}. 
Some of them may be tidally disrupted 
by the SMBH during the later orbit evolution 
under the resonant relaxation \citep{rauch_resonant_1996} 
in the captured star cluster \citep{perets_dynamical_2009}. 
We numerically evolve the angular momentum of a captured star ($L_{\rm{S}}$) 
in the range of 
$\pm L_{\rm{c}} (t_{\rm{S}}/T_{\rm{sRR}})^{1/2}$, 
where $t_{\rm{S}}$ is the time from its capture to the present and 
\begin{eqnarray}
    T_{\rm{sRR}}=\frac{1-e_{\rm{S}}^{2}}{e_{\rm{S}}^{2}}
    \left(\frac{M_{1}}{m_{\rm{S}}}\right)^{2}
    \frac{P_{\rm{S}}^{2}}{N(<a_{\rm{S}})t_{\omega}} 
\end{eqnarray}  
is the scalar resonant relaxation (sRR) timescale \citep{gurkan_resonant_2007}. 
Here $e_{\rm{S}}$, $m_{\rm{S}}$ and  $P_{\rm{S}}$ are, respectively, 
the eccentricity, mass, and orbital period of the captured star, 
$N(<a_{\rm{S}})$ is the number of stars with semimajor axis smaller than $a_{\rm{S}}$, 
and $t_{\omega}$ is the joint timescale of 
general relativistic (GR) and Newtonian (NT) precessions \citep{chen_there_2012}. 
A star is supposed to be tidally disrupted once its pericenter distance 
$r_{\rm{p}}\approx L_{\rm{S}}^{2}/2GM_{1}$ 
becomes smaller than the tidal disruption radius 
$r_{\rm{t}}=\eta (M_{1}/m_{\rm{S}})^{1/3}R_{\rm{S}}$ \citep{hills_possible_1975} 
with $R_{\rm{S}}=(m_{\rm{S}}/M_{\odot})^{0.56}R_{\odot}$ 
the stellar radius for $m_{\rm{S}} > 1M_{\odot}$ \citep{kippenhahn_stellar_2012}. 
We adopt $\eta =2$ to account for partial TDEs \citep[e.g.,][]{ryu_tidal_2020}.

\subsection{Estimation of the Model Parameters} \label{subsec:fitting} 

Our model has five free parameters to be estimated: 
(1) the SMBHB mass ratio $q_{\rm{MBHB}}$; 
(2) the SMBHB coalescence time $t_{\rm{m}}$; 
(3) the two-body relaxation timescale factor $\kappa$; 
(4) the decay timescale of the interaction rate 
for the young stellar populations of the NSC $\tau$; 
and (5) the scaling factor of the interaction rate $A_{\rm{e}}$. 
The first four parameters are estimated by conducting a large scale of grid searches 
in the four-dimensional parameter space from coarse to fine coarse, i.e., 
search the whole parameter space first 
and concentrate on the region where the fitting is better. 
The goodness of fitting is evaluated by jointly comparing 
the velocity and Galactocentric distance distributions of the mock HVS catalog 
in our model with the observation 
(see Appendix~\ref{appendix_sec:observation} and \ref{appendix_sec:fitting}). 
The last parameter $A_{\rm{e}}$ is determined by 
equating the expected number of halo HVSs by our model to 
that from the Multiple Mirror Telescope (MMT) HVS survey: 
$N_{\rm{HVS}}\approx 50$ \citep{brown_gaia_2018a}.

The orbit of the SMBHB may have a nonzero eccentricity $e_{\rm{MBHB}}$, 
which is unknown. 
Because $e_{\rm{MBHB}}$ evolves with time (see Section~(\ref{subsec:evolution})), 
we denote it with the value at the lookback time $t=500\, \rm{Myr}$: $e_{500}$. 
For the sake of computational expenses, 
we only explore our model for several typical values of $e_{500}$ 
from $0$ up to $0.75$ (see Table~\ref{table:results}). 
A larger eccentricity can be ruled out, 
as it would result in a rapid evolution of the SMBHB 
due to strong GW radiation at the pericenter 
and thus a burstlike HVS ejection that is 
inconsistent with the observation \citep{brown_mmt_2014}.

\subsection{Tidal Separations of Stellar Binaries from 
the Clockwise and Counterclockwise Disk} \label{subsec:CWD}

Young stellar binaries in the CWD formed at about $6\, \rm{Myr}$ ago \citep{paumard_two_2006} and
could also produce HVSs and captured stars. 
The HVSs originating in the CWD would not have enough flight time to reach the halo 
and are thus excluded from our HVS fitting. 
However, the captured stars from the CWD 
should still remain at the GC and be accounted for. 
We follow a similar procedure as introduced in Section~\ref{subsec:mock_catalog} 
to obtain a mock catalog of captured stars from the CWD, 
but with several modifications. 
First, we assume a top-heavy IMF $\rm{d}N/\rm{d}m\propto m^{-1.7}$ 
and place the simulated binary on a bound orbit 
at $0.05\, \rm{pc}<r<0.5\, \rm{pc}$ 
following the surface density $\Sigma (r)\propto r^{-2}$ based on the observation 
\citep{nayakshin_missing_2005,paumard_two_2006,bartko_extremely_2009,lu_stellar_2013}. 
The IMF implies that most of the early-type (O and B) binaries in the CWD 
should have B-type primaries, 
and the captured early-type stars should mainly consist of B-type stars 
(see S-CWD population in Section~\ref{subsec:results_Sstars}). 
Second, it is suggested that the dynamical evolution of the orbits in the CWD is 
mainly due to the eccentric disk instability rather than two-body scattering 
\citep{madigan_new_2009,generozov_hills_2020}, 
so we adopt a smaller critical semimajor axis $a_{\rm{c}}=2\, \rm{au}$ 
for the mock binary population 
and simulate their angular momentum evolution following \cite{generozov_constraints_2022}. 
Last, the interaction rate is modeled to 
decay exponentially with a timescale $\tau_{\rm{CWD}}$ 
after a burst formation at $6\, \rm{Myr}$ ago: 
\begin{align} \label{equ:t_escape_cwd}
    \Gamma_{\rm{CWD}}(t) & \propto  
    \rm{exp}\left(-\frac{6\, \rm{Myr}-t}{\tau_{\rm{CWD}}}\right), \ t\leq 6\, \rm{Myr}.
\end{align}  
The value of $\tau_{\rm{CWD}}$ can be estimated according to 
the number of O stars left in the CWD: $N_{\rm{O}}$, 
and that of the captured stars from the CWD: $N_{\rm{S-CWD}}$. 
Our model implies that their ratio should be 
\begin{equation} \label{equ:cal_rau_cwd}
    \frac{N_{\rm{O}}}{N_{\rm{S-CWD}}} = 
    \frac{f_{\rm{O}}}{(1-f_{\rm{TDE}})f_{\rm{b,CWD}}}
    \frac{\exp{(-\frac{6\, \rm{Myr}}{\tau_{\rm{CWD}}})}}
    {1-\exp{(-\frac{6\, \rm{Myr}}{\tau_{\rm{CWD}}})}},
\end{equation}
where $f_{\rm{TDE}}\approx 20\%$ is the fraction of the captured stars 
consumed by TDEs, 
$f_{\rm{b,CWD}}\approx 40\%$ is the refined binary fraction for stars 
$\geq 3.5M_{\odot}$ from the CWD, 
$f_{\rm{O}}\approx 14\%$ accounts for the different mass ranges 
for $N_{\rm{O}}$ ($\geq 16M_{\odot}$) and $N_{\rm{S-CWD}}$ ($\geq 3.5M_{\odot}$). 
Substituting $N_{\rm{O}}\approx 56$ \citep{fellenberg_young_2022} 
and $N_{\rm{S-CWD}}\approx 32$ (see Section~\ref{subsec:results_Sstars})
into Equation~(\ref{equ:cal_rau_cwd}) gives 
$\tau_{\rm{CWD}}\approx 26\, \rm{Myr}$.


\section{Results} \label{sec:results}

\subsection{Hypervelocity Stars in the Galactic Halo} \label{subsec:results_HVS}

\begin{figure*}
    \centering
    \includegraphics[width=\textwidth]{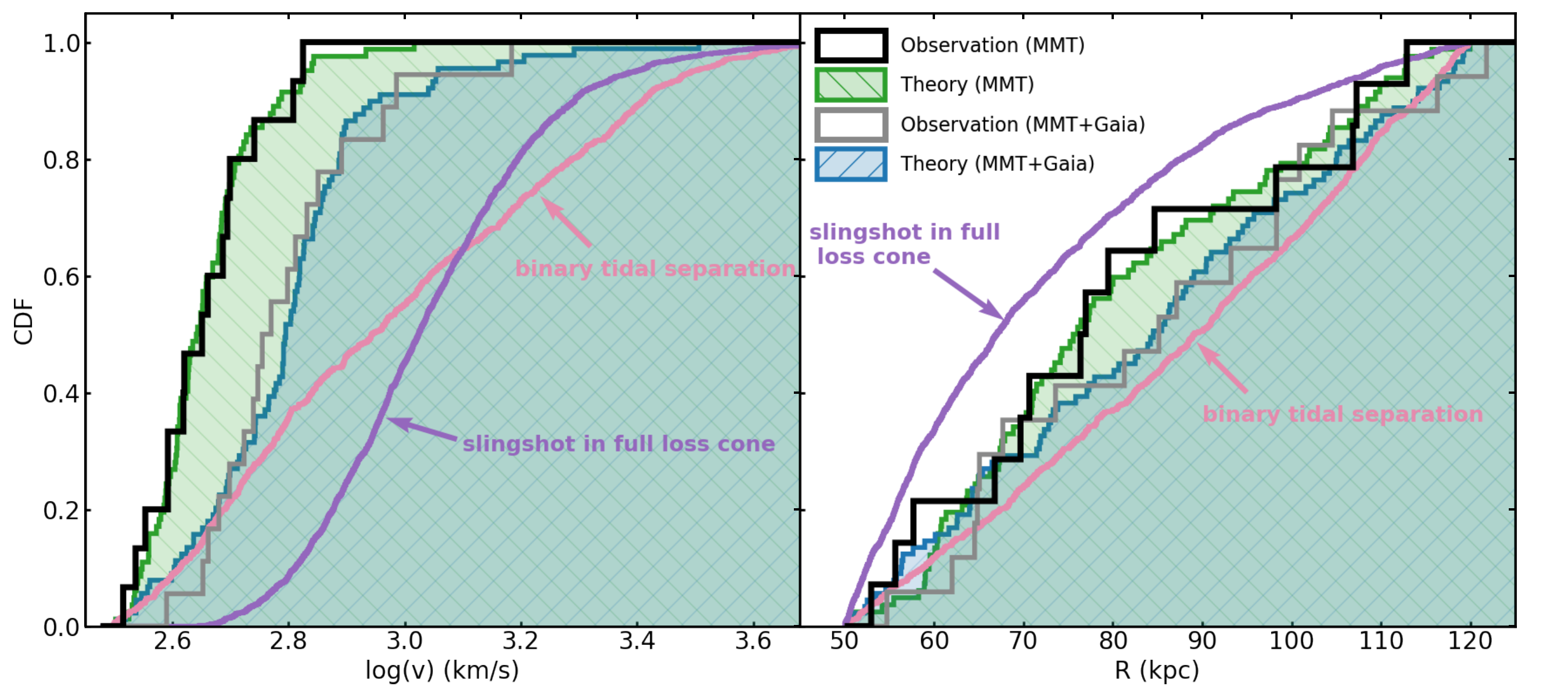}
    \caption{Cumulative distribution functions of velocity (left panel) and 
    Galactocentric distance (right panel) of the HVSs 
    of mass $\geq 2.5 M_\odot$ from our model 
    after propagation through the Galactic potential 
    to the distance between $50$ and $120\, {\rm kpc}$. 
    The results (back hatched green and hatched blue) are obtained with 
    the best-fitting parameters of our SMBHB model in the empty-loss-cone regime 
    to the observational data of the late B-type halo HVSs 
    without (\citealt{brown_gaia_2018a}; black) and 
    with (\citealt{kreuzer_hypervelocity_2020}; gray) 
    the corrections of the Gaia proper motions. 
    To illustrate the uniqueness and advantage of our model, 
    we also give the distributions obtained with 
    binary tidal separation by the SMBH Sgr~$\rm{A}^{*}$ (pink) and 
    the SMBHB slingshot in the full-loss-cone regime (purple).}
 \label{fig:hvs}
\end{figure*}

In Figure~\ref{fig:hvs}, 
we show the cumulative distribution functions of the velocity (\emph{v}-CDF) and 
Galactocentric distance (\emph{R}-CDF) of HVSs numerically expected 
by our SMBHB model in the empty-loss-cone regime.
In order to compare with the observation of the halo HVSs from the MMT survey, 
only the HVSs with mass $\geq 2.5 M_\odot$ and 
Galactocentric distances 50--120$\, \rm{kpc}$ are considered. 
We also show the numerical expectations 
of another two frequently discussed models, i.e., 
solely considering HVSs produced through 
binary tidal separations by a single SMBH (Model A) and 
slingshot ejections by an SMBHB in the full-loss-cone regime (Model B). 
Both of them are clearly incompatible with the observation 
as noticed by the previous studies 
\citep[e.g.,][]{sesana_hypervelocity_2007,zhang_spatial_2010,rossi_velocity_2014,
darbha_gravitational_2019,generozov_constraints_2022}. 
By contrast, 
our model can reproduce the observed \emph{v}-CDF and \emph{R}-CDF of HVSs well. 
In particular, only about 2.5\% of the halo HVSs, 
or half an HVS among the 15 halo HVSs in the MMT sample, 
are expected to have velocity above $700\, \rm{km\, s^{-1}}$ in our model 
(see $f_{>700}$ in Table \ref{table:results}), 
which is about 20 times smaller than the expectations of Model A and B. 
The deficit of HVSs above $700\ \rm{km\ s^{-1}}$ 
is unique and consistent with the observation.

The observed \emph{v}-CDF, in particular its upper limit, 
is associated with the orbit size and mass ratio of the SMBHB, 
and thus can be used to constrain our model parameters. 
We fit our model to both the observational data 
with \citep[MMT+Gaia sample;][]{kreuzer_hypervelocity_2020} and 
without \citep[MMT sample;][]{brown_gaia_2018a} the Gaia proper motion 
corrections (see Appendix~\ref{appendix_sec:fitting}). 
The best-fitting results are shown in Figure~\ref{fig:hvs} and 
the associated parameters are given in Table~\ref{table:results}. 
Because for the halo HVSs, 
the tangential velocities should not be significant \citep[e.g.,][]{yu_kinematics_2007} 
and the Gaia proper motion corrections
have large observational uncertainties \citep{kreuzer_hypervelocity_2020}, 
hereafter we mainly present the results based on the fitting to the MMT sample. 
Our best-fitting parameters 
(marked with $*$ in Table~\ref{table:results}) show that 
the SMBHB has a mass ratio of $2.9\times 10^{-3}$ 
and an eccentricity of $0.4$ at $500\, \rm{Myr}$ ago. 
The best-fitting two-body relaxation timescale is about 
twice (with the MMT sample) or one-third (with the MMT+Gaia sample) 
of the fiducial value at the SMBH influence radius 
(see $\kappa$ in Equation~(\ref{equ:T_r}) and Table~\ref{table:results}). 
Because both the Gaia proper motion corrections 
and the estimated two-body relaxation timescale are in large uncertainties, 
our results are consistent with the two-body relaxation process and 
suggest a weak effect of massive perturbers \citep[e.g.,][]{perets_massive_2007}.

\begin{deluxetable*}{clc|clrcc|rccc}
    \tablenum{1}
    \tablecaption{Best-fitting Parameters of the SMBHB Model in the Empty-loss-cone Regime 
    \label{table:results}}
    \tablewidth{0pt}
    \tablehead{
    Observation & \colhead{$e_{\rm{500}}$} & \colhead{$a_{\rm{500}}$} 
    & \colhead{$q_{\rm{MBHB}}$} & \colhead{$\kappa$} & \colhead{$\tau$} & \colhead{$f_{>700}$}
    & \colhead{$\sigma_{\rm{HVS}}^{2}$} & \colhead{$f_{\rm{HVB}}$} & \colhead{$a_{\rm{h}}$} 
    & \colhead{$e_{\rm{h}}$} & \colhead{$t_{\rm{h}}$}  \\
    sample &  & (au) & ($10^{-3}$)  &  & (Myr) & $(\%)$ & & $(\%)$ & (au) & & (Gyr)
    } 
    \startdata
         &  0.2  & 134.8  & 2.8 & 2.5  & 60  & 7.3 & 0.344  & 14.0 & 231.0 & 0.31 & 3.03 \\
    MMT  & 0.4$*$   & 156.7  & 2.9 & 2.1  & 40  & 2.5 & 0.257  & 13.3 & 239.2 & 0.54 & 1.51 \\
         & 0.5   & 239.7  & 3.2 & 2.7  & 20  & 3.1 & 0.262  & 9.0  & 263.9 & 0.56 & 0.87 \\
         & 0.75  & 299.4  & 4.4 & 1.0  & 43  & 8.6 & 0.528  & 14.7 & 362.4 & 0.78 & 0.67 \\
    \hline
         & 0.0  & 140.0  & 3.7 & 0.33 & 115 & 31 & 0.737 & 13.5 & 305.0 & 0.08 & 7.50 \\
         & 0.2  & 145.2  & 3.8 & 0.33 & 100 & 34 & 0.782 & 13.7 & 313.2 & 0.42 & 5.19 \\
    MMT  & 0.3  & 154.4  & 3.8 & 0.58 & 85  & 28 & 0.864 & 14.9 & 313.2 & 0.53 & 3.15 \\
    +    & 0.4  & 171.1  & 4.5 & 0.25 & 75  & 34 & 0.990 & 14.8 & 370.6 & 0.65 & 2.74 \\
    Gaia & 0.5  & 188.8  & 4.9 & 0.6  & 65  & 38 & 0.981 & 15.2 & 403.4 & 0.71 & 2.12 \\
         & 0.55 & 215.2  & 5.3 & 2.1  & 65  & 39 & 0.979 & 14.5 & 436.1 & 0.74 & 1.84 \\
         & 0.65 & 257.8  & 5.6 & 2.1  & 60  & 26 & 0.878 & 19.0 & 460.7 & 0.80 & 1.17 \\
         & 0.75 & 353.4  & 5.7 & 1.0  & 85  & 30 & 0.943 & 18.3 & 468.9 & 0.82 & 0.76 \\
    \enddata
    \tablecomments{
    The results are obtained for different eccentricities of the SMBHB at lookback time 
    $500\, {\rm Myr}$ ($e_{\rm 500}$). 
    MMT sample: 
    fitting to the observational data from the MMT survey of the halo HVSs \citep{brown_gaia_2018a}; 
    MMT+Gaia sample: 
    fitting to the observational data from the MMT survey with the 
    Gaia proper motion corrections \citep{kreuzer_hypervelocity_2020};   
    $a_{500}$: the SMBHB semimajor axis at lookback time $500\, {\rm Myr}$;  
    $q_{\rm{MBHB}}$: the mass ratio of the SMBHB; 
    $\kappa$: the scaling factor of the two-body relaxation timescale 
    relative to the fiducial estimate for the NSC (see Equation~(\ref{equ:T_r})); 
    $\tau$: the decay timescale of the interaction rate for young stars of the NSC; 
    $f_{>700}$: the fraction of HVSs with velocities above $700\, \rm{km\, s^{-1}}$; 
    $\sigma_{\rm{HVS}}^{2}$: the parameter measuring the goodness of fitting 
    (see Equation~(\ref{equ:goodness})); 
    $f_{\rm{HVB}}$: the fraction of HVB; 
    $a_{\rm{h}}$, $e_{\rm{h}}$, and $t_{\rm{h}}$: 
    the semimajor axis (hardening radius), eccentricity, and the lookback time 
    when the SMBHB became hard, respectively. }
\end{deluxetable*}

After the dynamical interaction with the SMBHB, 
a loose binary with $r_{\rm{bt}}\gtrsim a_{\rm{MBHB}}$ would be tidally separated, 
and a tight binary with $r_{\rm{bt}}\ll a_{\rm{MBHB}}$ would survive 
and be ejected intactly possibly as a hypervelocity binary (HVB). 
Our results ($f_{\rm{HVB}}$ in Table~\ref{table:results}) suggest that 
about $10\%$ of (one or two) halo HVSs are HVBs, 
consistent with the discovery of HVS3 as the only candidate
of HVBs or rejuvenated hypervelocity blue stragglers 
\citep{edelmann_he_2005,perets_runaway_2009}.

\begin{figure*}
    \centering
    \includegraphics[width=0.8\textwidth]{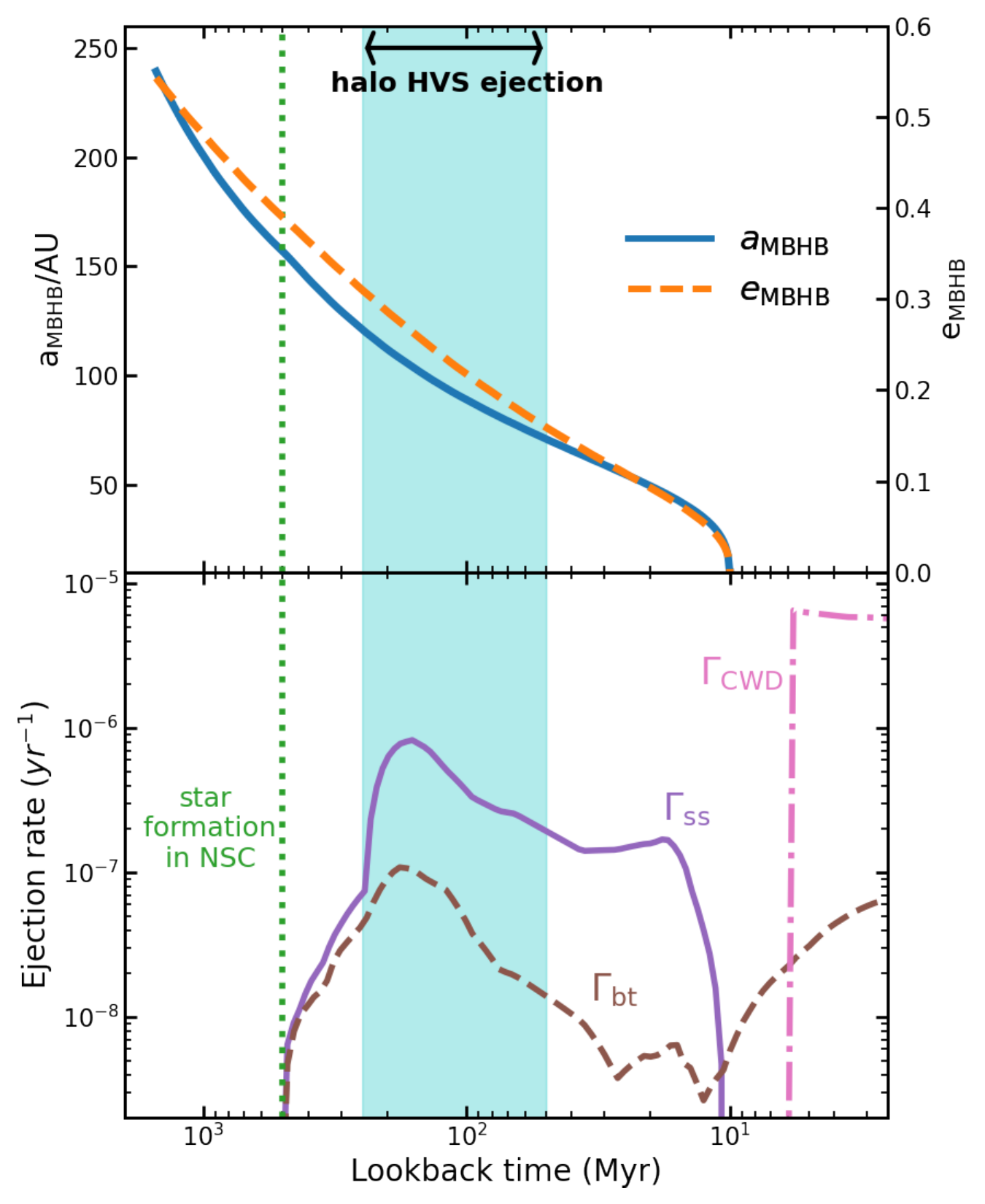}
    \caption{
    The evolution of the semimajor axis ($a_{\rm{MBHB}}$) 
    and eccentricity ($e_{\rm{MBHB}}$) of the SMBHB 
    from the hardening radius $a_{\rm h}$ (upper panel) and 
    the production rate of HVSs of mass $\geq 2.5 M_\odot$ (lower panel)
    as functions of lookback time 
    in our model with best-fitting parameters.    
    The solid purple and dashed brown curves are, respectively, 
    the production rate of HVSs through 
    the slingshot effect ($\Gamma_{\rm ss}$) and 
    the binary tidal separation ($\Gamma_{\rm bt}$). 
    The dashed-dotted pink curve ($\Gamma_{\rm CWD}$) shows the HVS production rate by the 
    tidal separations of stellar binaries from the CWD.
    The vertical green dotted line marks the beginning of the formation of 
    the 500$\, $Myr population in the NSC. 
    The cyan shadow region shows the time window when 
    the halo HVSs observed in the MMT survey are ejected. 
    }
    \label{fig:evolution}
\end{figure*}

In Figure~\ref{fig:evolution}, 
we show the variation of the production rate of early-type 
($\geq 2.5M_{\odot}$) HVSs with the lookback time. 
The production rate increases with the star formation of the 500$\, $Myr population and 
reaches a maximum of about $10^{-6}\, \rm{yr}^{-1}$ at $150\, \rm{Myr}$ ago. 
Thereafter, the production rate decreases with time 
following the dynamical relaxation of the newly formed stars 
and the evolution of the SMBHB. 
The HVSs are mainly produced through slingshot ejections 
until about $12\, {\rm Myr}$ ago, 
when the SMBHB is about to coalesce. 
After the SMBHB coalescence, 
all the HVSs are produced through binary tidal separations. 
During the past $6\, \rm{Myr}$, 
the tidal separations of binaries from the CWD give rise to 
a burst ejection of HVSs with a peak rate 
$\Gamma_{\rm{CWD}}\simeq 7 \times 10^{-6}\, \rm{yr}^{-1}$.

\subsection{Captured Stars at the Galactic Center} \label{subsec:results_Sstars}

\begin{figure}
    \centering
    \includegraphics[width=0.85\textwidth]{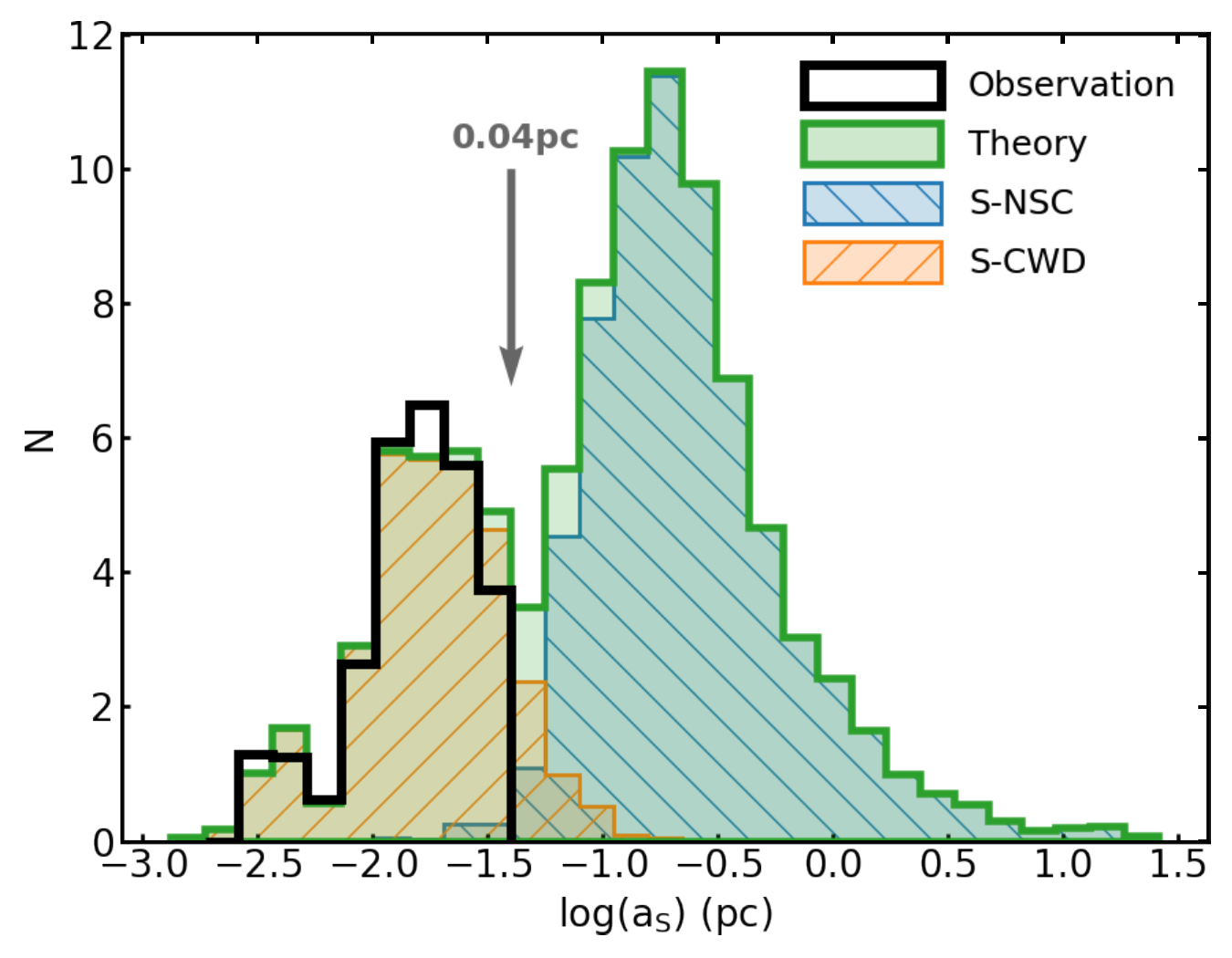}
    \caption{Semimajor axis distribution of the captured stars after the 
    tidal separations of binaries predicted by our SMBHB model (green), 
    compared to the observed distribution of the semimajor axis of the S-stars with the
    selection-effect truncation at $0.04\, {\rm pc}$ (gray arrow) from 
    \cite{gillessen_update_2017} and 
    \citeauthor{peissker_s62_2020a}(2020; black).
    The captured stars expected by our SMBHB model consist of two populations: 
    S-NSC population (blue back hatched) with progenitor binaries from the NSC and 
    captured before the SMBHB coalescence,
    and the S-CWD population (orange hatched) with progenitor binaries 
    from the CWD and predominantly captured after the SMBHB coalescence.}
    \label{fig:Sstar}
\end{figure}

After the tidal separations, 
the counterparts of the ejected stars are captured by the SMBH Sgr~$\rm{A}^{*}$ 
on highly eccentric orbits. 
The young S-stars observed at the GC 
\citep{gillessen_update_2017,peissker_s62_2020a} 
are most likely the remnants of the tidal separations 
\citep{gould_sagittarius_2003,zhang_galactic_2013}. 
In Figure~\ref{fig:Sstar}, 
we show the semimajor axis ($a_{\rm{S}}$) distribution of the captured stars 
of mass $\geq 3.5 M_\odot$ (spectral types earlier than B5V) 
expected by our model. 
The captured stars can be grouped into two populations based on their origins: 
one from the 500$\, $Myr and 80$\, $Myr populations of the NSC (S-NSC population) 
and the other from the CWD (S-CWD population). 
The S-NSC population consists of 67 late B-type stars, 
inferred by equating the number of the halo HVSs expected by our model to 
that from the MMT HVS survey: $N_{\rm{HVS}}\approx 50$ \citep{brown_gaia_2018a}. 
They are predominantly captured before the SMBHB coalescence and distributed at 
$0.05\, \rm{pc}(1.3^{\prime\prime}) \lesssim a_{\rm{S}} \lesssim 3\, \rm{pc}(1.3^{\prime})$ 
with a peak at about $0.2\, \rm{pc}(5^{\prime\prime})$. 
Our model expectations for the S-NSC population are largely consistent with 
the F2 and F3 features observed in the GC that 
consist of 73 young stars with median projected distances 
$\sim 6^{\prime\prime}$ -- $8^{\prime\prime}$ and eccentricity 0.7 
\citep{fellenberg_young_2022}. 
Meanwhile, the S-CWD population is captured after the SMBHB coalescence 
and mostly has $a_{\rm{S}}\leq 0.04\, {\rm pc}$. 
It consists of 31 B-type stars and one O-type star ($> 20 M_\odot$), 
inferred by equating the number of captured stars of $a_{\rm{S}}\leq 0.04\, \rm{pc}$ 
expected by our model to the observed value: $N_{\rm{S}}\approx 27$ 
(see Appendix~\ref{appendix_sec:observation}). 
Overall, the semimajor axis distribution of the captured stars 
is expected to be double peaked 
with one peak at $0.018\, \rm{pc}$ and the other at $0.18\, \rm{pc}$. 
Note that the numbers of captured stars given above are rounded and 
subject to the Poisson fluctuation of $N_{\rm{HVS}}$ and $N_{\rm{S}}$. 

We find that about $20\%$ of captured stars are consumed by TDEs, 
most of which belong to the S-CWD population.  
The TDE consumption leads to a deficiency of captured stars at 
$a_{\rm{S}}\approx 1300\, \rm{au}$, 
because the GR and NT precessions cancel out (they have opposite directions), 
and the sRR effect is the most efficient therein. 
The local deficiency is well consistent with the observed distribution of S-stars.

Figure~\ref{fig:Sstar} shows that the observed semimajor axis distributions 
of the S-stars at $\leq 0.04\, \rm{pc}$ 
(see Appendix~\ref{appendix_sec:observation}) can be well reproduced 
as long as they are predominantly made up of the S-CWD population 
and basically uncontaminated by the S-NSC population. 
This scenario can only be achieved by our SMBHB model, 
as the capture of the S-NSC population within 0.04 pc is blocked 
by the secondary IMBH, 
just like the ejections of the halo HVSs above 
$700\, \rm{km\, s^{-1}}$. 
Our SMBHB model thus suggests that nearly all the S-stars 
should originate in the CWD and be spectral B-type stars. 
The recent measurements of eight S-stars (including S2) reveal that 
they have a mass of 8--14$M_\odot$ and ages 
$\lesssim 14\, \rm{Myr}$ \citep{habibi_twelve_2017}, 
consistent with our expectations.

\subsection{A Past Supermassive Black Hole-–Intermediate-mass Black Hole Binary 
at the Galactic Center} \label{subsec:result_SMBHB}

We have shown above that 
the existence of a past SMBHB at the GC is suggested by 
the velocity distribution and in particular 
the deficit of the halo HVSs above $700\, \rm{km\, s^{-1}}$ 
and the semimajor axis distribution and ages of S-stars. 
Because the halo HVSs were ejected from the GC $50$--$250\, \rm{Myr}$ ago and 
S-stars were captured in the past few million years, 
the orbital evolution of the SMBHB in the past $250\, \rm{Myr}$ 
can be well constrained. 
Given that the SMBHB evolves mainly due to 
the GW radiation and the slingshot effect, 
its orbit can be integrated backward to the hardening radius $a_{\rm{h}}$ 
(see Equation~(\ref{equ:SMBHB_decay})). 
Figure~\ref{fig:evolution} shows 
the evolution of the SMBHB in our model. 
The SMBHB became hard about $1.5\, \rm{Gyr}$ ago 
when $a_{\rm{MBHB}}\approx 240\, \rm{au}$ and $e_{\rm{MBHB}}\approx 0.54$,
ejected the halo HVSs 
when $a_{\rm{MBHB}}\approx $ 80--120$\, \rm{au}$, 
and eventually coalesced about $10 \, {\rm Myr}$ ago.

The large eccentricity at the hardening time implies that 
the SMBHB most probably formed after the merger of the MW with a dwarf galaxy 
on an orbit of large eccentricity \citep[e.g.,][]{gualandris_eccentricity_2022}. 
The most likely candidate is the 
Gaia-Sausage-Enceladus (GSE) dwarf galaxy, 
which had a total stellar mass of $M_* \simeq 5\times 10^8 M_\odot$ and 
merged with the MW about $10\, {\rm Gyr}$ ago on a highly eccentric orbit 
\citep{helmi_merger_2018,naidu_reconstructing_2021}. 
If the GSE dwarf galaxy harbors an IMBH at the center, 
the empirical relationship of the SMBH and galactic total stellar masses 
for late-type galaxies \citep{greene_intermediatemass_2020} 
gives a mass of $M_{\rm IMBH}\simeq 0.7\times 10^4 M_\odot$, 
consistent with the result obtained with our SMBHB model: 
$M_{\rm IMBH}\simeq 1.5\times 10^4 M_\odot$.

The coalescence of the SMBHB would lead to 
a gravitational recoil of the SMBH Sgr~$\rm{A}^{*}$ \citep{campanelli_maximum_2007}. 
With the SMBHB mass ratio $\simeq 4\times 10^{-3}$, 
the orbit parameters of the GSE dwarf galaxy \citep{naidu_reconstructing_2021}, 
and the dimensionless spin $0.94 (0.5)$ of the SMBH Sgr~$\rm{A}^{*}$ 
obtained with the Event Horizon Telescope (EHT) \citep{collaboration_first_2022}, 
the recoil velocity is about $0.5(0.3)\, \rm{km\, s^{-1}}$. 
This is consistent with the observed proper motion of Sgr~$\rm{A}^{*}$: 
$-0.58\pm 2.23\, \rm{km\, s^{-1}}$ and $-0.85\pm 0.75\, \rm{km\, s^{-1}}$ 
in the directions parallel and vertical to the Galactic plane, 
respectively \citep{reid_proper_2020},  
which is larger than the Brownian motion velocity: 
$0.2\, \rm{km\, s^{-1}}$ \citep{merritt_brownian_2007}.

\section{Conclusion and Discussion} \label{sec:dis}

HVSs currently observed in the Galactic halo were 
ejected from the GC 50--250$\, \rm{Myr}$ ago 
and thus carry rich information about the dynamical environment of the GC 
in the recent past, 
especially that associated with the SMBH Sgr~$\rm{A}^{*}$. 
In this Letter, we propose a new model for producing the late B-type halo HVSs 
by considering the young stars/binaries of the NSC interacting with 
an evolving SMBHB in the empty-loss-cone regime. 
With extensive numerical simulations, 
we show that the observed velocity and Galactocentric distance distribution 
of the halo HVSs can be well reproduced by our model, 
and the unique upper truncation of the velocity distribution 
at about $700\, \rm{km\, s^{-1}}$ implies a past SMBHB at the GC. 
By fitting our model expectation to the observation of HVSs, 
we find that the SMBHB has a mass ratio of $\simeq 3\times 10^{-3}$ 
and coalesced about $10\, \rm{Myr}$ ago. 
Our results suggest that the S-stars observed in the innermost arcsecond 
around Sgr~$\rm{A}^{*}$ 
are from tidally separated binaries originating in the CWD 
after the SMBHB coalescence.

\begin{figure}
    \centering
    \includegraphics[width=0.85\textwidth]{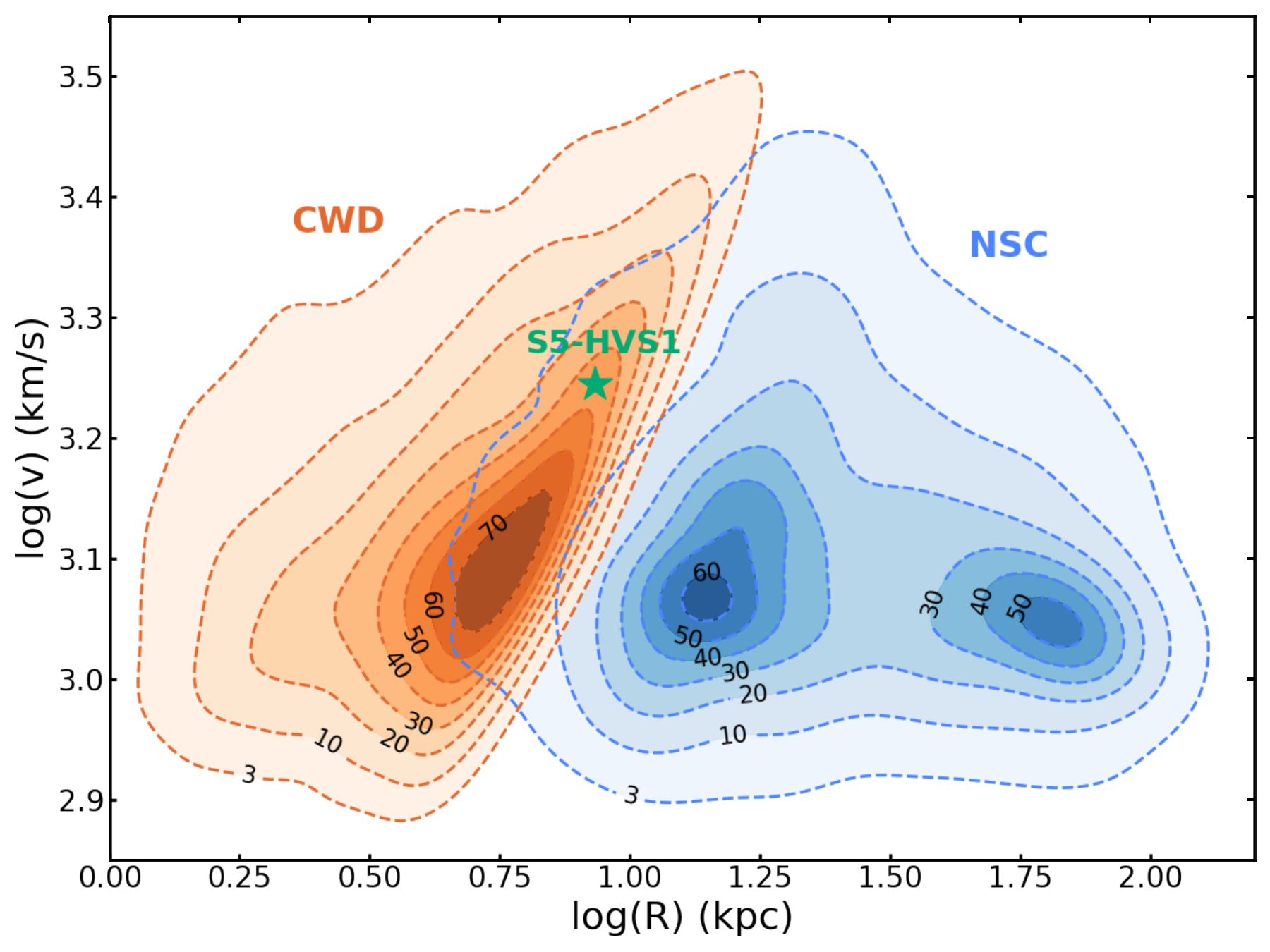}
    \caption{All-sky detection probability map of the HVSs of mass 
    $\geq 1.5M_{\odot}$ and Galactocentric distance $\geq 1\, {\rm kpc}$ 
    predicted for the $S^{5}$ survey. 
    The contours are labeled with the HVS number per unit logarithmic area, i.e.,  
    $\rm{d}^2{N}/\rm{d}\log{(R)}\rm{d}\log{(v)}$. 
    The detection probability for the expected HVSs with progenitors 
    from the CWD is in orange, 
    and that from the young stellar population of the NSC is in blue. 
    The HVS S5-HVS1 is marked with a green ``$\star$,'' 
    which is located close to the peak detection probability of the HVSs
    originating from the CWD. }
    \label{fig:des}
\end{figure}

We explore the detectability of HVSs by some surveys 
following the procedure outlined in previous studies 
\citep{marchetti_predicting_2018,evans_constraints_2023}, 
assuming that they are ejected from the GC isotropically. 
The assumption of isotropic ejections is definitely not accurate, 
especially for HVSs from the CWD, 
so our predictions for HVS surveys here are only tentative. 
According to our model, 
about 30 HVSs originating in the CWD were ejected in the past $6\, \rm{Myr}$, 
most of which are B and A stars.  
We find that the expected number of these HVSs that are visible in 
Gaia Data Release 3 \citep{vallenari_gaia_2022,katz_gaia_2023} 
is only of order $10^{-3}$. 
This is consistent with the Gaia's non-detection of HVSs so far. 
Considering that an HVS S5-HVS1 of mass $2.35\, M_\odot$ 
was recently discovered 
with an extremely high velocity of $1755 \, \rm{km\ s^{-1}}$ 
at Galactocentric distance $8.5\, \rm{kpc}$ \citep{koposov_discovery_2020}
from the Southern Stellar Stream Spectroscopy Survey 
\citep[$S^{5}$;][]{li_southern_2019}, 
we also calculate the expectations of HVSs for the $S^{5}$ survey. 
Figure~\ref{fig:des} shows the expected detection probability distribution of the HVSs 
of mass $\geq 1.5M_{\odot}$ on a logarithmic velocity-distance diagram. 
The HVSs originating in the young stellar population of the NSC and 
those originating in the CWD are clearly separated on the diagram: 
the former are concentrated around $v\approx 1300\, \rm{km\ s^{-1}}$ and 
$R\approx 6\, {\rm kpc}$, 
while the latter have a typical velocity of $1200\, \rm{km\ s^{-1}}$ and 
their Galactocentric distance distribution shows two peaks 
at about $15$ and $70\, {\rm kpc}$. 
The double-peaked distance distribution 
is due to the requirement of the heliocentric velocity above $800\, \rm{km\ s^{-1}}$ 
according to the HVS inspection strategy of the $S^5$ survey \citep{koposov_discovery_2020}. 
The position of S5-HVS1 on the diagram 
and the much higher (about 10 times) detection probability  
suggest that it is most likely produced through the tidal separation of a binary 
from the CWD rather than that from the NSC. 
A CWD origin for S5-HVS1 has been speculated previously 
\citep[e.g.,][]{koposov_discovery_2020,generozov_hills_2020,lu_former_2021}. 
However, it faces the difficulty that the estimated age of S5-HVS1 
\citep[25--93$\, \rm{Myr}$;][]{koposov_discovery_2020} is greater than that of the CWD. 
Considering the uncertainties of age estimates for individual stars 
\citep[e.g.,][]{soderblom_ages_2010} 
and of the astrophysical parameters of S5-HVS1
(e.g., its effective temperature, metallicity and surface gravity, 
which are crucial to the age estimate) 
inferred by several works 
\citep[e.g.,][]{andrae_gaia_2023,creevey_gaia_2023,zhang_parameters_2023}, 
the more accurate determination of the S5-HVS1's age is needed.

We notice that a large fraction of early-type stars are born 
in triples or higher-order multiples 
\citep[e.g.,][]{chini_spectroscopic_2012,sana_southern_2014}. 
To explore the impact of triples, 
we assume that $50\%$ of binaries are in stable hierarchical triple systems 
\citep{moe_mind_2017}. 
We generate a mock catalog of triples by adopting
the stable criterion in \cite{mardling_tidal_2001} 
and drawing the parameters of the inner and outer binaries 
from the same distributions introduced in Section~\ref{subsec:mock_catalog} 
following \cite{toonen_evolution_2020}. 
We simulate the evolution of the mock triples 
in our model of SMBHB in the empty-loss-cone regime. 
We find that during the slow orbital diffusion, 
most (about $86\%$) of the triples would become unstable 
as the eccentricities of their outer binaries are tidally excited by the SMBH
\citep[e.g.,][]{zhang_spatial_2010}. 
For those stable triples, 
only $27\%$ of the tidal separations of their outer binaries 
will end up with unbound ejections. 
Therefore, the contribution of triples to the total HVS/HVB population 
is negligible ($\lesssim 2\%$).

The scaling parameter $A_{\rm{e}}$ 
in our model accounts for the magnitude of the HVS production rate.
It turns out to be several tens to over 100 
($A_{\rm{e}}=141.34$ for our best-fitting result), 
suggesting a strongly enhanced HVS production rate during 50--250$\, {\rm Myr}$ ago 
compared to the estimate given by the empty-loss-cone theory 
\citep[e.g.,][]{perets_massive_2007}. 
Our preliminary investigations suggest that the enhancement could be 
caused by the disklike distribution of the young NSC populations. 
In this scenario, young stars of the NSC are born in a gaseous disk that is 
similar to the circumnuclear disk observed in the GC 
\citep{becklin_farinfrared_1982,genzel_neutralgas_1985,
guesten_aperture_1987,jackson_neutral_1993} and form a stellar disk. 
Some stars would escape from the stellar disk azimuthally 
due to the eccentric disk instability \citep{madigan_new_2009} 
and get deposited at the loss-cone boundary, 
which is more efficient than the two-body scattering and boosts the HVS ejection rate. 
We would emphasize that this will not change the loss-cone regime, 
because once escaped from the dense environment of its birthplace, 
the orbital evolution of a star around the loss cone 
is still driven by the two-body scattering of the predominant old field stars. 
The disklike origin is supported by the observed anisotropic distribution of HVSs \citep{brown_mmt_2012} 
and has already been investigated in some previous studies 
\citep[e.g.,][]{lu_spatial_2010,zhang_spatial_2010,zhang_galactic_2013}. 
However, the detailed studies about the enhanced rate and the morphology of 
young NSC populations are beyond the scope of this study, 
and we plan to address them in a following work.

\begin{acknowledgments}
We thank the referee for comments and suggestions 
that help to improve the quality of our work. 
This work is supported by 
the National Natural Science Foundation of China (NSFC Nos. 11721303 and 11991053), 
China Manned Space Project with No. CMS-CSST-2021-A06, 
and National Key R\&D Program of China (grant No. 2020YFC2201400). 
S.L. acknowledges the support of the National Science Foundation of China 
under grant NSFC No. 12473017. 
K.W. acknowledges support from the National Natural Science Foundation of China 
(12041305, 12033005), 
the Tianchi Talent Program of Xinjiang Uygur Autonomous Region, 
and the China--Chile Joint Research Fund (CCJRF No. 2211). 
\end{acknowledgments}

\software{
NumPy \citep{harris2020array}, 
SciPy \citep{2020SciPy-NMeth}, 
astropy \citep{2022ApJ...935..167A}, 
pandas \citep{mckinney-proc-scipy-2010}, 
Matplotlib \citep{Hunter:2007}, 
IPython \citep{PER-GRA:2007}, 
Gala \citep{adrian_price_whelan_2020_4159870}. 
}

\appendix
\restartappendixnumbering

\section{Hardening of Supermassive Black Hole Binary in the Empty-loss-cone Regime} \label{appendix_sec:harden_rate}
\begin{figure*}
    \centering
    \includegraphics[width=0.5\textwidth]{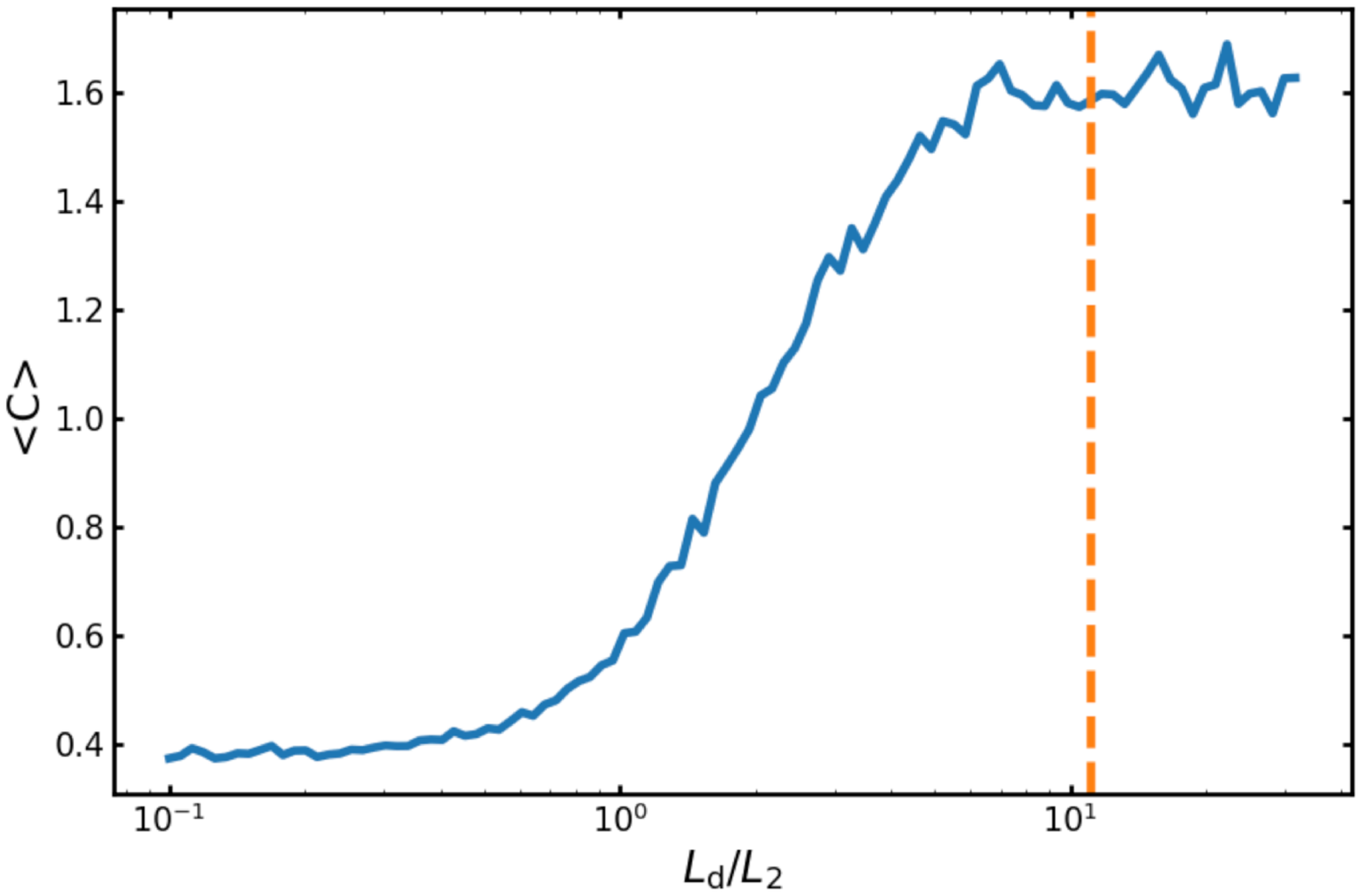}
    \caption{Mean energy exchange fraction $\langle C \rangle$ as a function of 
    the diffusion step length the angular momentum $L_{\rm{d}}$ in a unit of 
    $L_{2}=(T_{\rm{d}}/T_{\rm{r}})^{1/2}L_{\rm{c}}$. 
    The orange vertical line marks the boundary between 
    the full and empty-loss-cone regimes.} 
    \label{fig:getC}
  \end{figure*}

The evolution of an SMBHB driven by slingshot ejections of ambient stars 
can be described by two parameters $\langle C \rangle$ and $\langle B \rangle$, 
defined as, respectively, 
the averaged fractional exchange of specific energy and angular momentum
during each slingshot ejection \citep{quinlan_dynamical_1996}. 
It is suggested that $\langle C \rangle$ is a constant 
for a hard SMBHB in the full-loss-cone regime 
\citep{quinlan_dynamical_1996,sesana_interaction_2006}. 
However, its value in the empty-loss-cone regime should be a function of $L_{\rm{d}}$.  
We evaluate $\langle C \rangle$ for different $L_{\rm{d}}$ 
with scattering experiments following the procedure in \cite{sesana_interaction_2006} 
and present the result in Figure~\ref{fig:getC}. 
We find that $\langle C \rangle$ generally increases with $L_{\rm{d}}$ 
and converges to the same value as that in the full-loss-cone regime
\citep[$\approx 1.6$;][]{sesana_interaction_2006} when $L_{\rm{d}}$ is large enough. 
As for the parameter $\langle B \rangle$, 
it is related to $\langle C \rangle$ as 
\begin{equation} \label{equ:B_C}
    \frac{\langle B \rangle}{\langle C \rangle} = 
    \left[\frac{GM_{1}}{a_{\rm{MBHB}}^{3}(1-e_{\rm{MBHB}}^{2})}\right]^{1/2}
    \frac{\Delta L_{\rm{MBHB}}}{\Delta E_{\rm{MBHB}}}.
\end{equation}
Assuming that the ratio 
$\Delta L_{\rm{MBHB}}/\Delta E_{\rm{MBHB}}$ does not change significantly
with $e_{\rm{MBHB}}$ and calibrating Equation~(\ref{equ:B_C}) to 
$\langle B \rangle/\langle C \rangle=1$ when $e_{\rm{MBHB}}=0$
\citep{sesana_interaction_2006}, we have 
\begin{equation} \label{equ:B_C_fit}
    \langle B \rangle     \simeq  
    (1-e_{\rm{MBHB}}^{2})^{-1/2}\langle C \rangle .
\end{equation}
 
\section{Scattering Experiments} \label{appendix_sec:scattering}

To obtain the outcome of 
the gravitational encounter between a SMBHB and a stellar binary, 
we conduct extensive (about $3\times 10^{7}$) scattering experiments 
with program \texttt{Fewbody} \citep{fregeau_stellar_2004} 
covering the full parameter space of initial conditions. 
We consider a stellar binary of mass $m_{1}=m_{2}=3M_{\odot}$ approaching 
an SMBHB with the primary mass of $M_{1}=4\times 10^{6}M_{\odot}$ on a parabolic orbit. 
For the sake of computational efficiency, 
both the SMBHB and the stellar binary have zero initial eccentricities, 
which also eliminate the need of specifying the arguments of the pericenter. 
Then the initial condition is defined by eight parameters: 
(1) The mass ratio of the SMBHB: $q_{\rm{MBHB}}$; 
(2-3) The semimajor axis of the SMBHB and the stellar binary: 
$a_{\rm{MBHB}}$ and $a_{*}$; 
(4) The pericenter distance of the stellar binary's orbit around the SMBHB: $r_{\rm{p}}$; 
(5-8) The orientations of the SMBHB and the stellar binary relative to 
the stellar binary's orbit around the SMBHB, 
described by the inclination ($\theta_{\rm{MBHB}}$ and $\theta_{*}$) 
and the longitudes of ascending node ($\Omega_{\rm{MBHB}}$ and $\Omega_{*}$). 
For each point in the parameter space, 
we perform 100 scattering experiments with uniformly sampled mean anomalies 
between $0$ and $2\pi$ for both the SMBHB and the stellar binary. 
From the results, 
we obtain the probabilities of different outcomes for the stellar binary, 
the distribution of the energy gain from the slingshot effect or 
the binary tidal separation, 
and the distribution of $\zeta \equiv a'_{*}/a_{*}$ for a survived binary with 
$a'_{*}$ the postinteraction semimajor axis to account for the tidal heating effect.

Given a specific set of initial conditions, 
we obtain the outcome of the interaction by interpolating the 
probabilities of different outcomes 
with \emph{LinearNDInterpolator} in scipy.interpolate module 
and sampling the energy gain and $\zeta$ 
from the weighted mixture of the distribution 
at 256 most adjacent points in the 8-dimensional parameter space, 
where the weights are evaluated based on the logarithmic distances for 
($a_{\rm{MBHB}},a_{\rm{*}},r_{\rm{p}}$) and the linear distances for other parameters.

\section{The Observation Sample of Hypervelocity Stars and S-stars} 
\label{appendix_sec:observation}

We use two observation samples of HVSs for the model fitting 
(see Table~\ref{table:hvs_mmt}). 
The MMT sample is compiled by selecting HVSs with mass $\geq 2.5M_{\odot}$ 
and Galactocentric distances $50\, \rm{kpc}\leq R \leq 120\, \rm{kpc}$ 
from the MMT HVS survey \cite[Table 2 in][]{brown_gaia_2018a}. 
We remove four of them (HVS7, HVS8, HVS12, and HVS17) that 
are suggested to have non-GC origins 
\citep{brown_gaia_2018a,irrgang_hypervelocity_2018} 
and keep HVS5 and HVS6 whose origins are controversial 
\citep{brown_gaia_2018a,irrgang_hypervelocity_2018,
kreuzer_hypervelocity_2020,irrgang_blue_2021}. 
We have tested that excluding HVS5 and HVS6 would not have much impact on our results, 
as our model parameters are mainly determined by 
the upper limit of velocity distribution (see Equation~(\ref{equ:goodness})).  
The final MMT sample consists of 15 HVSs. 
The MMT+Gaia sample 
is compiled in the same way based on Table A.1 in \cite{kreuzer_hypervelocity_2020},  
which accounts for proper motions from 
the second data release of Gaia \cite[Gaia DR2;][]{brown_gaia_2018}. 
Compared to the MMT sample, the MMT+Gaia sample includes 
another three candidates (B129, B329, B458), 
but HVS1 is discarded as it becomes a clear outlier in terms of the Galactocentric distance. 

\begin{deluxetable*}{lDDDDc}
    \tablenum{A1}
    \tablecaption{The Observational Sample of HVSs. 
    \label{table:hvs_mmt}}
    \tablewidth{0pt}
    \tablehead{
    Object & \twocolhead{$v^{\rm{M}}$} & \twocolhead{$R^{\rm{M}}$} & 
    \twocolhead{$v^{\rm{G}}$} & \twocolhead{$R^{\rm{G}}$} & \colhead{References} \\ 
           & \twocolhead{($\rm{km\ s^{-1}}$)}  & \twocolhead{(kpc)} & 
    \twocolhead{($\rm{km\ s^{-1}}$)}  & \twocolhead{(kpc)}  &   
    } 
    \decimalcolnumbers
    \startdata
    HVS1$^{\rm{M}}$    & $669.8\pm 6.6$    & $106.9\pm 15.3$  &  710  &  166    &          \\
    HVS4$^{\rm{M,G}}$  & $551.7\pm 7.3$    & $69.7\pm 10.5$   &  630  &  81     &          \\
    HVS9$^{\rm{M,G}}$  & $458.8\pm 6.1$    & $77.0\pm 12.2$   &  710  &  116    &          \\
    HVS10$^{\rm{M,G}}$ & $417.0\pm 4.6$    & $53.0\pm 5.9$    &  500  &  74     &          \\
    HVS13$^{\rm{M,G}}$ & $418.5\pm 10.8$   & $107.3\pm 19.6$  &  780  &  122    &          \\
    HVS15$^{\rm{M,G}}$ & $328.5\pm 8.1$    & $66.8\pm 9.7$    &  450  &  62     &          \\
    HVS16$^{\rm{M,G}}$ & $344.6\pm 7.3$    & $70.7\pm 11.6$   &  680  &  65     &          \\
    HVS18$^{\rm{M,G}}$ & $449.0\pm 8.5$    & $79.5\pm 11.1$   &  560  &  98     &          \\
    HVS19$^{\rm{M,G}}$ & $496.2\pm 13.1$   & $98.3\pm 15.3$   &  920  &  87     &          \\
    HVS20$^{\rm{M,G}}$ & $392.1\pm 8.7$    & $76.4\pm 10.9$   &  970  &  105    &          \\
    HVS21$^{\rm{M,G}}$ & $391.9\pm 7.5$    & $112.9\pm 21.7$  &  580  &  101    &          \\
    HVS22$^{\rm{M,G}}$ & $487.4\pm 11.5$   & $84.7\pm 13.5$   &  1530 &  98     &          \\
    HVS24$^{\rm{M,G}}$ & $358.6\pm 7.6$    & $55.7\pm 7.7$    &  460  &  65     &          \\
    HVS5$^{\rm{M,G}}$  & $644.0\pm 7.5$    & $49.8\pm 5.8$   &  650   &  43     & (a, b, c)    \\
    HVS6$^{\rm{M,G}}$  & $501.1\pm 6.3$    & $57.7\pm 7.2$   &  530   &  55     & (c)        \\
    B129$^{\rm{G}}$    &        .          &       .          &  390  &  93     &          \\
    B329$^{\rm{G}}$    &        .          &       .          &  480  &  68     &          \\
    B458$^{\rm{G}}$    &        .          &       .          &  570  &  85     &          \\
    \hline
    HVS7  & $397.7\pm 6.8$    & $53.2\pm 6.5$   &  500   &  49     & (a, b, c)    \\
    HVS8  & $413.3\pm 2.6$    & $58.3\pm 10.8$  &  500   &  42     & (a, b, c)    \\
    HVS12 & $417.4\pm 8.1$    & $67.2\pm 8.7$   &  570   &  77     & (a, b)      \\
    HVS17 & $439.5\pm 4.6$    & $48.7.8\pm 4.3$ &  460   &  35     & (b, c)      \\
    B481  &        .          &       .         &  460   &  47     & (c)        \\
    B576  &        .          &       .         &  680   &  51     & (b)        \\
    B1085 &        .          &       .         &  500   &  43     & (b)        \\
    \enddata\
    \tablecomments{
    The HVSs with mass $\geq 2.5M_{\odot}$ and Galactocentric distance 
    $50\ \rm{kpc}\leq R \leq 120\ \rm{kpc}$. 
    $v^{\rm{M}}$ and $R^{\rm{M}}$ are the original data of 
    velocities and Galactocentric distances from Table 2 in \cite{brown_gaia_2018a}. 
    $v^{\rm{G}}$ and $R^{\rm{G}}$ are the data 
    after the Gaia proper motion corrections from 
    Table A.1 in \cite{kreuzer_hypervelocity_2020}. 
    The objects with a ``M'' (``G'') superscript are compiled as 
    the MMT (MMT+Gaia) sample and used for our model fitting. 
    We list the references in 
    determining the HVS origin other than \cite{brown_gaia_2018a}. 
    \bf{Reference.} 
    (a) \cite{irrgang_hypervelocity_2018};
    (b) \cite{kreuzer_hypervelocity_2020}; 
    (c) \cite{irrgang_blue_2021}.
     } 
\end{deluxetable*}

The observation sample of S-stars is compiled by selecting early-type stars 
($\geq 3.5M_{\odot}$) from \cite{gillessen_update_2017} and \cite{peissker_s62_2020a}. 
Among them, only 30 stars have measured orbital parameters, 
which are distributed isotropically around Sgr~$\rm{A}^{*}$
with thermally distributed eccentricities \citep{gillessen_monitoring_2009}. 
For those with undetermined orbital parameters, 
we assume that they follow the same distribution 
and express their semimajor axes $a$ 
as posterior probability distribution functions $P(\rm{a})$. 
We calculate $P(\rm{a})$ with Bayes formula: 
$P(a\mid R_{\rm{p}})\propto P(R_{\rm{p}}\mid a)P_{0}(a)$, 
where $R_{\rm{p}}$ is the projected distance to Sgr~$\rm{A}^{*}$ 
measured in \cite{gillessen_update_2017} 
and $P_{0}(a)$ is the prior distribution assumed to be uniform. 
The final sample consists of about 27 S-stars with semimajor axis $\leq 0.04\ \rm{pc}$.

\section{Fitting to the observation} \label{appendix_sec:fitting}

Given a set of the four parameters (see Table~\ref{table:para_range}), 
we obtain a mock catalog of HVSs by 
repeating the procedure in Section~\ref{subsec:mock_catalog} 
until the number of mock HVSs is larger than 150. 
We fit the \emph{v}-CDF and \emph{R}-CDF 
of the mock HVS catalog to the observation.  
In practice, we divide $v$ between $[300, 2000]\, \rm{km\, s^{-1}}$ and 
$R$ between $[50, 120]\, \rm{kpc}$ into 500 uniform bins. 
At each bin of index $i$, 
we evaluate the \emph{v}-CDF and \emph{R}-CDF of our mock HVS catalog 
as $X(v_{\rm{i}})$ and $Y(R_{\rm{i}})$ 
and those of the observational HVS sample as 
$X_{0}(v_{\rm{i}})$ and $Y_{0}(R_{\rm{i}})$. 
We find the best-fitting parameters by minimizing 
\begin{eqnarray} \label{equ:goodness}
    \sigma_{\mathrm{HVS}}^{2}= \eta_{\rm{v}} \sum_{i} 
    \frac{\left[X(v_{\rm{i}})-X_{0}(v_{\rm{i}})\right]^{2}}
    {X_{0}^{-1}(v_{\rm{i}})} 
    +\eta_{\rm{R}}\sum_{i} 
    \frac{\left[Y(R_{\rm{i}})-Y_{0}(R_{\rm{i}})\right]^{2}}{Y_{0}^{-1}(R_{\rm{i}})}, 
\end{eqnarray}
where the denominators account for the statistical error, 
and $\eta_{\rm{v}}$ and $\eta_{\rm{R}}$ are the weights of the two terms.  
We set $\eta_{\rm{v}} = 1$ and $\eta_{\rm{R}} = 1/8$ 
when fitting the MMT sample, 
as the velocity is measured about 8 times more precise than the distance 
in terms of the averaged relative error \citep{brown_gaia_2018a}. 
As for the MMT+Gaia sample, 
we set $\eta_{\rm{v}}=\eta_{\rm{R}} = 1/8$ considering that 
both velocity and distance measurements have large uncertainties 
\citep{kreuzer_hypervelocity_2020}. 
By construction, our fitting method is more sensitive to the high-velocity 
($v\gtrsim 700\, \rm{km\, s^{-1}}$) tail of the velocity distribution. 

\begin{deluxetable*}{lcc}
    \tablenum{A2}
    \tablecaption{The Parameter Space Explored in the Model Fitting. 
    \label{table:para_range}}
    \tablewidth{0pt}
    \tablehead{
    Parameter & Total Range & Fine-coarse Range 
    } 
    \startdata
    $q_{\rm{MBHB}}$     & $[10^{-3},\ 10^{-1}]$  & $[2\times 10^{-3},\ 10^{-2}]$   \\
    $t_{\rm{m}}$ (Myr)  & $[1,\ 50]$         & $[5,\ 20]$              \\
    $\kappa$            & $[0.01,\ 10]$      & $[0.1,\ 5]$            \\
    $\tau$ (Myr)        & $[1,\ 500]$        & $[10,\ 150]$            \\
    \enddata
    \tablecomments{
    ``Total Range'' is the complete parameter space explored in this study 
    and ``Fine-coarse Range'' shows the parameter space where the fitting is better.} 
\end{deluxetable*}


\end{document}